\documentclass{article}

\usepackage{microtype}
\usepackage{graphicx}
\usepackage{amsfonts}
\usepackage{subfigure}
\usepackage{siunitx}
\usepackage{multirow}
\usepackage{amsmath}
\usepackage{tikz}
\usetikzlibrary{shapes.geometric, arrows, positioning}
\usepackage{float}
\restylefloat{table}
\usepackage{booktabs} 

\usepackage{hyperref}

\newcommand\CC[1]{{\mathcal{#1}}}
\newcommand\BB[1]{{\mathbb{#1}}}
\newcommand\B[1]{{\mathbf{#1}}}

\usepackage[accepted]{icml2019}

\icmltitlerunning{GeNet: Deep Representations for Metagenomics}

\begin{document}

\twocolumn[
  \icmltitle{GeNet: Deep Representations for Metagenomics}



\icmlsetsymbol{equal}{*}
\icmlsetsymbol{sharp}{$\sharp$}

\begin{icmlauthorlist}
\icmlauthor{Mateo Rojas-Carulla}{cam,mpi}
\icmlauthor{Ilya Tolstikhin}{equal,goo}
\icmlauthor{Guillermo Luque}{mpibio}
\icmlauthor{Nicholas Youngblut}{mpibio}
\icmlauthor{Ruth Ley}{mpibio}
\icmlauthor{Bernhard Sch{\"o}lkopf}{mpi}

\end{icmlauthorlist}

\icmlaffiliation{cam}{Department of Engineering, University of Cambridge, Cambridge, England}
\icmlaffiliation{goo}{Google AI, Z{\"u}rich, Switzerland}
\icmlaffiliation{mpi}{Max Planck Institute for Intelligent Systems, T{\"u}bingen, Germany}
\icmlaffiliation{mpibio}{Max Planck Institute for Developmental Biology, T{\"u}bingen, Germany}

\icmlcorrespondingauthor{Mateo Rojas-Carulla}{mrojascarulla@gmail.com}


\vskip 0.3in
]



\printAffiliationsAndNotice{\icmlEqualContribution} 

\begin{abstract}
We introduce GeNet, a method for shotgun metagenomic classification from raw DNA 
sequences that exploits the known hierarchical structure between labels for training. 
We provide a comparison with state-of-the-art methods Kraken and Centrifuge on datasets 
obtained from several sequencing technologies, in which dataset shift occurs. 
We show that GeNet obtains competitive precision and good recall, 
with orders of magnitude less memory requirements. Moreover, we show 
that a linear model trained on top of representations learned by GeNet 
achieves recall comparable to state-of-the-art methods on the aforementioned datasets, 
and achieves over $90\%$ accuracy in a challenging pathogen detection problem. 
This provides evidence of 
the usefulness of the representations learned by GeNet for downstream biological tasks.
\end{abstract}

\section{Introduction}\label{sec:intro}

The last two decades have seen an exponential decrease in the cost of 
next generation DNA sequencing, transforming the field of microbiome science~\citep{turnbaugh2007human, pasolli2019extensive}. 
A \emph{microbiota} is a community of microorganisms residing in a multi-cellular organism, and the 
\emph{microbiome} is the collective genetic material of this microbiota. 
Recently, mechanisms 
by which the human microbiota has an effect on 
a variety of health outcomes have been discovered~\citep{sonnenburg2016diet}, and its 
responsiveness to dietary and lifestyle interventions~\citep{walker2011dominant, wu2011linking} 
may lead to effective ways to prevent disease and improve health outcomes. 
Given a biological sample, studying the effect of the microbiota on its host  
requires as a first step understanding \emph{which} 
microorganisms it contains. 
Nonetheless, the output of sequencing technologies are DNA \emph{reads}, noisy 
substrings of the genomes present in the biological sample. 
How does one process such a collection of reads to understand 
which organisms are present, and in which amounts? 
The problem of \emph{shotgun metagenomic classification} aims 
to assign to each read the corresponding host organism, 
and is 
an essential first step before downstream analysis can be carried-out. 
Currently, state-of-the-art methods such as 
Centrifuge~\citep{kim2016centrifuge} 
and Kraken~\citep{wood2014kraken} rely on sequence alignment, 
and match each read against a 
large database of known genomes. This requires high amounts of 
memory for storing such databases, and becomes challenging as the amount of noise 
in the reads increases. 

The availability of more affordable sequencing technologies has been accompanied 
by noisier reads. Oxford Nanopore's MinION~\citep{jain2016oxford} 
comes with error rates close to $10\%$, orders of magnitude higher than typical noise 
levels for Illumina, a more expensive technology. 
Sequence alignment based methods suffer from increasing 
ambiguity as noise increases. Machine learning systems, 
on the other hand, can \emph{learn} from the noise distribution of the input 
reads. Moreover, a classification model learns a \emph{mapping} from input 
read to class probabilities, and thus does not require a database 
at run-time. Finally, machine learning systems provide 
\emph{representations} of DNA sequences which can be leveraged 
for downstream tasks. 

\paragraph{Contributions.}
In this paper, we introduce GeNet, a convolutional neural network model 
for shotgun metagenomic classification. GeNet is trained end-to-end from 
\emph{raw} DNA sequences and exploits a hierarchical taxonomy between 
organisms using a novel architecture. 
We compare GeNet against state-of-the-art methods on 
real datasets, and show that GeNet achieves similar precision and 
good recall, despite the occurrence of strong dataset shift. We 
show in two examples that the representations of DNA sequences learned 
by GeNet can be successfully exploited by a linear classifier, achieving 
over $90\%$ recall at the species level in the introduced datasets and 
over $90\%$ accuracy in a challenging pathogen detection problem, 
outperforming baseline features. 
Finally, the trained weights only require $126$MB of 
memory which is orders of magnitude smaller than databases for 
current metagenomic classifiers, providing advantages for portable, 
affordable technologies targeted for field use. 

\section{Problem statement and model}\label{sec:model}
\subsection{Metagenomic classification}

\begin{figure*}
\tikzstyle{arrow} = [thick,->,>=stealth]
\tikzstyle{taxonode} = [rectangle, rounded corners, minimum width=2cm, minimum height=1cm,text centered, draw=black, fill=blue!10, align=center]
\begin{tikzpicture}[node distance=2.5cm]
\node (phylum) [taxonode] {\texttt{Phylum}\\$N_1=38$};
\node (class) [taxonode, right of=phylum] {\texttt{Class}\\$N_2 = 68$};
\node (order) [taxonode, right of=class] {\texttt{Order}\\$N_3 = 150$};
\node (family) [taxonode, right of=order] {\texttt{Family}\\$N_4 = 318$};
\node (genus) [taxonode, right of=family] {\texttt{Genus}\\$N_5 = 791$};
\node (species) [taxonode, right of=genus] {\texttt{Species}\\$N_6 = 1870$};
\node (leaf) [taxonode, right of=species] {\texttt{Leaf}\\$N_7 = 3375$};

\draw [arrow] (phylum) -- (class);
\draw [arrow] (class) -- (order);
\draw [arrow] (order) -- (family);
\draw [arrow] (family) -- (genus);
\draw [arrow] (genus) -- (species);
\draw [arrow] (species) -- (leaf);

\end{tikzpicture}
\caption{Taxonomic ranks in the tree $\CC{T}$ used to train GeNet, and the number of taxa (nodes) 
at each level. The coarser level considered is \emph{Phylum}, and the finer level before the leaves 
is \emph{Species}. A taxonomic tree containing all living organisms has significantly 
more taxa.}\label{fig:taxotree}
\end{figure*}
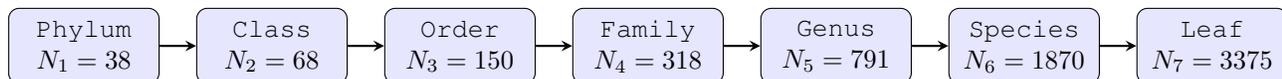

  We consider the problem of \emph{shotgun metagenomic classification} 
  from raw DNA sequences. We are given a collection of organisms $G_1, \ldots, G_K$ for which the complete genome is known\footnote{We use the terms organism and genome interchangeably for the rest of the paper.}. Each genome is a string consisting of four \emph{nucleotides} \texttt{A}, \texttt{C}, \texttt{T} and \texttt{G}. 
  Moreover, a hierarchical taxonomy $\CC{T}$ encoding the similarities between 
  the $K$ genomes is available. 
  $\CC{T}$ is a tree, and each of the $K$ genomes lies in one of its leaves. 
  There are $L$ levels in the tree corresponding to different 
  \emph{taxonomic ranks}, ranking from coarser (e.g., \emph{Life}) to finer 
  (e.g., \emph{Species}). Each level $\ell$ contains $N_{\ell}$ nodes, also called 
  \emph{taxa} (\emph{taxon} in singular). The taxonomic ranks used for training GeNet and 
  the corresponding number of taxa are depicted in 
  Figure~\ref{fig:taxotree}, and represent only a fraction of the known biological taxonomy. 
  As an illustration, humans and chimpanzees are Homonidae, and thus both belong 
  to the same taxonomic rank \emph{Family}. Nonetheless, they both belong to a different 
  \emph{Genus}, and thus to a different \emph{Species}, two finer taxonomic ranks.
  
  When sequencing DNA from a biological sample, a specific \emph{sequencing technology} is used. 
  Some widely used technologies include Illumina, Pacbio and Nanopore. The output 
  of these technologies are called \emph{reads} and are noisy substrings $\B{s}$ of the genomes 
  present in the sample. 
  The length of the reads and the noise they contain is 
  characteristic to the sequencing technology, and results in nucleotides being 
  flipped, deleted or added. The distribution 
  of this noise, which we call \emph{technology specific noise}, is in practice empirically estimated~\citep{mcelroy2012gemsim}. 
  Illumina technologies produce short reads (around $100$ 
  nucleotides), all of the same length, while Nanopore technologies produce longer reads of varying length (roughly 
  between $1,000$ and $10,000$ nucleotides).
  
    Given an input read $\B{s}$ which is a noisy substring of genome $G_k$, 
  $\CC{T}$ defines a unique labeling $\B{y} = (y_1, \ldots, y_L)$, where $y_{\ell}\in\{1,\ldots, N_{\ell}\}$ 
  is the correct label, or taxon, at the $\ell$-th level of $\CC{T}$. 
  The goal of \textbf{metagenomic classification} is to predict the correct taxon for a read 
  $\B{s}$ at different levels in $\CC{T}$. For the example in Figure~\ref{fig:taxotree}, metagenomic classification at the 
  Phylum level consists of finding the correct taxon for read $\B{s}$ among $N_1 = 38$ possibilities.

  If one is able to correctly label a read to the corresponding host genome $G_k$, the correct hierarchical labelling vector $\B{y}$ can be directly deduced by tracing the unique path to the root in $\CC{T}$. 
  However, it may not be possible to correctly classify a sequence at 
  the finer levels of the tree. 
  Genomes close in the taxonomy may share significant portions 
  of their DNA, and reads from such genomes lead to ambiguous classification, especially 
  at the finer levels of the taxonomic tree~\citep{jain2018high}. Classification of short reads 
  is also often ambiguous, and is still desirable to classify at coarser levels 
  in the tree. 
  Of particular interest when analysing biological properties of a sample 
  are the levels of \emph{genus} and \emph{species}. For example, the study of pathogenesis (the biological mechanisms 
  leading to disease) requires  
  species level classification. 

\subsection{GeNet: a convolutional model for metagenomic classification}\label{sec:genet}

We propose GeNet, a model for metagenomic classification based on convolutional neural networks. 
GeNet is trained end-to-end from raw DNA sequences using standard backpropagation with cross-entropy loss. 
Given an input sequence $\B{s}\in\{0, 1, 2, 3\}^d$, we first extract features $\B{h} = G(\B{s}) \in \BB{R}^q$ 
using a Resnet like model~\citep{he2016deep} and a fully connected layer. 
This representation $\B{h}$ is then mapped to $L$ softmax layers, each of size $N_{\ell}$, 
which provides a probability distribution over the $N_{\ell}$ possible taxa at each level $\ell$ of 
the taxonomy $\CC{T}$. The number of known genomes is significantly higher than could be easily handled by 
standard neural network architectures, since it would result in hundreds of thousands of classes. We 
train GeNet with a representative subset of all the known genomes, details are given in Section~\ref{sec:exp}.

Denote by $\{W_{\ell}\}_{\ell=1}^L$ the weight matrices mapping $\B{h}$ 
to softmax vectors of size $N_{\ell}$, 
the number of possible taxa at the $\ell$-th level of the tree, 
where $W_{\ell} \in \BB{R}^{q \times N_{\ell}}$. 
We allow for predictions at level $\ell - 1$ to inform predictions at level $\ell$. 
For example, knowing that a read is likely to come from a bacterial organism narrows down 
the possible taxa at finer levels in $\CC{T}$. 
To that end, we include in our model matrices  
$\{U_{\ell}\}_{\ell=1}^{L}$ which allow that the unnormalised probability vector over taxa at level 
$\ell - 1$ contributes to the unnormalised probability vector at 
level $\ell$, where $U_{\ell} \in \BB{R}^{N_{\ell -1} \times N_{\ell}}$ and $U_0$ is a zero matrix. 
This leads to the following 
unnormalised probability vector over labels at each level of $\CC{T}$: 
\begin{equation}\label{eq:softmax}
\hat{\B{y}}_{\ell} = 
\texttt{ReLU}(W_{\ell}\B{h}) + \texttt{ReLU}(U_{\ell}\hat{\B{y}}_{\ell - 1}).
\end{equation}

A diagram of GeNet is provided in Appendix~\ref{sec:traindetails}. 

\paragraph{Training pipeline}
GeNet receives as input a noise parameter $p$, or \emph{base-calling probability}, and the size 
$r_{max}$ of input reads. We choose to train with uniform noise, that is, each nucleotide in a read 
is flipped with probability $p$, and is replaced uniformly among the remaining three nucleotides. 
This allows us to remain agnostic to technology specific noise, and can be partially 
corrected after training if necessary as described in Section~\ref{sec:da}. 
During training, the genomes $G_1, \ldots, G_K$ are stored in memory and new mini-batches 
are generated on-the-fly. For each iteration of stochastic gradient descent, we produce a new mini-batch 
of size $M$ by 
randomly selecting $M$ genomes among the $K$ available, selecting a random location in each and extracting a read of 
varying length from this location. 
We add uniform noise with parameter $p$ to each read, and pad it with zeroes 
to fit into the fixed input length $r_{max}$. 
For reads longer than the input length, only the first $r_{max}$ letters of longer 
reads are considered\footnote{Further work could try to split a longer read in smaller pieces, classify each 
shorter read, and aggregate the results into a classification decision.}. This allows us to classify reads of varying length. 
Uniform noise was chosen to remain agnostic to technology specific noise. 
The procedure for sampling a mini-batch is described in Algorithm~\ref{alg:sample_mb}. 
    
\paragraph{Loss function}
Let $\hat{\B{y}}_{\ell} \in \BB{R}^{N_{\ell}}$ be the unnormalised probability 
vector predicted by the network over the taxa at level $\ell$ as defined in 
Equation~\ref{eq:softmax}, and 
let $\B{y} = (y_1, \ldots, y_L)$ be the true taxa for 
the input sequence $\B{s}$ at each level of the taxonomy. 
We denote by $\B{y}_{\ell}$ the one-hot encoding 
of $y_{\ell}$, so that $\B{y}_{\ell}$ is a zero vector of size $N_{\ell}$ 
except for location $y_{\ell}$, which equals one. 

At each level of $\CC{T}$, we compute the cross-entropy 
loss 
$$
C_{\ell}(\hat{\B{y}}_{\ell}, \B{y}_{\ell}) = -\sum_{k = 1}^{N_{\ell}} \B{y}_{\ell, k} \log\left(\hat{\B{y}}_{\ell, k}\right).
$$

The distribution of taxa in $\CC{T}$ gives rise to class imbalance 
which we found necessary to correct  
for successful training. To illustrate, assume that $K$ genomes 
are seen uniformly during training. 
Consider now a coarser level in the tree, such as \emph{Domain} (not used to train GeNet), 
which contains 
the taxa \emph{Bacteria}, \emph{Virus} and \emph{Archaea}. For this illustration, 
assume that of the $K$ genomes, $90\%$ are bacteria, $8\%$ virus and $2\%$ archaea, meaning 
that the taxa at the \emph{Domain} level are highly imbalanced. Similar imbalances are 
likely to appear at most levels of the tree. If this imbalance is not corrected, any non bacterial 
organism wrongly classified as a bacteria is likely to be wrongly classified at finer levels 
of $\CC{T}$.
Therefore, each level of the tree uses a vector 
$\B{v}_{\ell} \in \BB{R}^{N_{\ell}}$ which down-weights the contribution to the loss of more abundant taxa. 
In the previous example, $\B{v}_{Domain} = (1/0.9, 1/0.08, 1/0.02)$. 
If a uniform distribution is assumed at the leaf level, the resulting $L$ weight vectors 
are determined by the $K$ genomes and the tree $\CC{T}$. 

The overall loss is then the weighted sum of cross-entropy losses at all the levels 
in the tree, 
\begin{equation}
C(\hat{\B{y}}, \B{y}) = \sum_{\ell=1}^L \B{v}_{\ell, y_\ell} C_{\ell}(\hat{\B{y}}_{\ell}, \B{y}_{\ell}). \nonumber
\end{equation}
GeNet is trained using standard stochastic 
gradient descent with Nesterov momentum~\citep{sutskever2013importance}.

\paragraph{Architecture choice}
Recurrent neural networks are the standard tool when considering 
sequential inputs. Nonetheless, we chose a convolutional architecture mainly 
due to i) speed constraints and ii) more success achieving high validation accuracies. 
We found that achieving good coverage of the $K$ genomes 
is important to achieve high validation accuracies, i.e., 
we must observe most parts of all the genomes. 
The combined length of the genomes used to train GeNet is over $10$ billion and 
GeNet trained for about one week on a NVIDIA P-100 GPU until convergence. 
The input to the network is a sum of a one-hot encoding of the letters in the sequence, 
an embedding of the sequence and a positional embedding as proposed in~\citet{convseq2seq17}. 
Both the embedding matrix and positional embedding matrix are learned during training. 
Many applications involving sequential data have benefited from convolutional 
architectures, for example machine translation~\citep{convseq2seq17}, text classification~\citep{conneau17} 
and video classification~\citep{karpathy2014large}.

\subsection{Domain adaptation and general purpose representations}\label{sec:da}
Since any test 
dataset obtained from real sequencing technologies represents a probability distribution 
that differs from the training distribution of GeNet, we face dataset shift. 
First, the noise distribution of real reads is not uniform, and depends 
on the sequencing technology. Moreover, the proportion of genomes is often not uniform, 
since longer genomes are over-represented, and the sample contains genomes in different 
proportions altogether. 
Genomes which were not observed during training may also be present. 
A priori, there are therefore no guarantees regarding the generalization performance 
of GeNet to unseen data if no further training takes place. Experiments in Section~\ref{sec:exp_da} report generalization 
performance of GeNet on real datasets. 

Nonetheless, intermediate activations of the network provide representations 
of the DNA reads which can be used for downstream tasks. 
Given a supervised learning problem with training data $\CC{D} = \{(\B{s}_i, z_i)\}_{i=1}^n$, where $\B{s}$ are DNA 
sequences and $z$ are outcomes we wish to predict from these sequences, we propose to use GeNet to compute representations of the training inputs $\B{h}_i = G(\B{s}_i)$, 
where $\B{h}_i$ is the last hidden layer of GeNet, see Section~\ref{sec:genet}. We then train a classification 
model on the transformed dataset $\widetilde{\CC{D}} = \{(\B{h}_i, z_i)\}_{i=1}^n$ and evaluate test performance.

We show that the representations in the last hidden layer of GeNet can be used for downstream tasks 
in Section~\ref{sec:exp_features} in two different examples. 
First, 
we consider datasets of Nanopore reads, for which dataset shift occurs, and 
we show that a linear model trained on GeNet representations  
achieves classification accuracy competitive 
with the state-of-the-art. 
Second, we consider the binary classification problem of deciding whether a 
read comes from a pathogenic organism, several of which were not observed during training, and show 
that a linear model achieves over $90\%$ test accuracy.
Such use of pre-trained features
has had remarkable success in image recognition (see \citet{rawat2017deep} and 
references therein) and natural language processing with representations such as
word2vec~\citep{mikolov2013distributed}. It is reassuring that the same holds true 
for metagenomics. 

\begin{algorithm}[tb]
   \caption{\texttt{SAMPLE MINI-BATCH}}
   \label{alg:sample_mb}
\begin{algorithmic}
   \STATE {\bfseries Input:} Reference genomes $G_1, \ldots, G_K$. Taxonomic tree $\CC{T}$. 
   Input size $r_{max}$, minimum read length $r_{min}$, base-calling error probability $p$. Mini-batch size $M$.
   
   \STATE {\bfseries Returns:} Mini-batch of noisy sequences $S$ and corresponding hierarchical 
   labels $Y$. 
   
   \STATE Initialize $S = $\texttt{[]}, $Y = $\texttt{[]}.
   \FOR{$j=1$ {\bfseries to} $M$}
   \STATE Select a \emph{genome} $G_k$ uniformly at random.
   
   \STATE Select a \emph{location} $j\in\{1, \ldots, L_k - r_{max}\}$ uniformly at random, where $L_k$ is the length of $G_k$. 
   \STATE Select a \emph{read length} $r\in\{r_{min}, \ldots, r_{max}\}$ uniformly at random. 
   \STATE Define the read $\B{s}_j = G_k(i : i + r)$, add uniform flipping noise with probability $p$ and 
    pad with zeroes at the end of the read, so that $\B{s}_j$ has length $r_{max}$. 
    \STATE Obtain the corresponding \emph{label vector} $\B{y}_j$ from $\CC{T}$.
    \STATE $S.\mathrm{add}(\B{s}_j)$, $Y.\mathrm{add}(\B{y}_j)$
   \ENDFOR
   
\end{algorithmic}
\end{algorithm}

\section{Related work}\label{sec:related}

State-of-the-art methods for metagenomic classification rely on sequence alignment. 
They use a large database of known genomes $G_1, \ldots, G_K$ and given a 
read $\B{s}$, do an exhaustive search to find one or more genomes $G_k$ such that $\B{s}$ 
is a substring of $G_k$. Since reads are often noisy, and the genomes in the sample may not 
exactly match any genome in the reference database, a notion of distance  
is used to compare strings. BLAST~\citep{altschul1990basic} performs this exhaustive search, 
but does not scale to real world datasets with millions of reads. State-of-the-art 
methods compress the database of genomes and strike a balance between accuracy and speed. 
These methods include Kraken~\citep{wood2014kraken} and Centrifuge~\citep{kim2016centrifuge}. 

The machine learning perspective on metagenomic classification is not new.~\citet{Busia353474} 
build a convolutional network for metagenomic classification of short reads (under $200$ nucleotides) 
on 16S data, and achieve good performance 
on a series of datasets. These data come from very specific parts of a conserved gene 
only present in prokaryotic organisms, excluding viruses and eukaryotes. 
Our approach focuses on the shotgun 
setting and supports reads originating from 
any part of the genome.
 Moreover,~\citet{Busia353474} do not analyse the re-usability of 
features learned by the network for downstream tasks.~\citet{nissen2018binning} introduce a method for 
the related problem of metagenomic binning using Variational Autoencoeders (VAE)~\citep{kingma2013auto}, 
and show that the representations learned by the VAE are useful for clustering. Their system is trained 
on co-abundance and composition data, not raw DNA sequences. To our knowledge, GeNet is the first system 
trained from raw DNA sequences for shotgun metagenomics. Finally,~\citet{feng2018hierarchical} introduce 
a deep learning system exploiting function hierarchy for gene function prediction. 

In machine learning, using label hierarchy to boost classification performance is not new. 
\citet{silla2011survey} provide a survey of standard methods used by practitioners 
in different fields. One option is to flatten the tree, essentially performing 
classification at the leaf node and disregarding the taxonomy. The availability of the 
tree does however allow for hierarchical classification during evaluation, since a correctly 
classified leaf is correct at all levels in the taxonomy. Another approach is having local 
classifiers at different nodes in the tree~\citep{vural2004hier,cerri2014hierarchical}, 
often of limited applicability for hierarchies with a large number of nodes. 
In metagenomic classification, such an approach would suffer from short read lengths for which ambiguity 
is high, since the likelihood of a sequence belonging to several unrelated genomes becomes higher. 
\citet{babbar2013flat} 
provide data-dependent bounds indicating when exploiting the hierarchy helps compared to  
using the flattened tree.~\citet{levatic2015importance} also analyse 
the usefulness of exploiting label hierarchy in a wide range of classification tasks. 
Recent papers have encoded the hierarchy in a neural network architecture.~\citet{zhu2017b} propose a multi-branch convolutional network, 
in which each branch aims to predict the label at a specific level of the tree similarly to 
GeNet. However, contrary to this approach, GeNet predicts labels at all levels of the tree from the 
same hidden representation simultaneously. 

\section{Experiments}\label{sec:exp}
\begin{table*}
\centering
  \begin{tabular}{lSSSSSSSS}
    \toprule
    \multirow{2}{*}{\textbf{Method}} &
      \multicolumn{2}{c}{$\CC{D}_1$ (\%)} &
      \multicolumn{2}{c}{$\CC{D}_2$ (\%)} &
      \multicolumn{2}{c}{$\CC{D}_3$ (\%)} & 
      \multicolumn{2}{c}{$\CC{D}_4$ (\%) } \\
      & {Genus} & {Species} & {Genus} & {Species} & {Genus} & {Species} & {Genus} & {Species}\\
      \midrule
    GeNet & ${94.6}$ & ${96.1}$ & ${93.9}$ & ${95.5}$ & ${98.5}$ & ${98.8}$ & ${98.5}$ & ${98.8}$ \\
    GeNet top 5 &  ${97.7}$ & ${98.2}$ & ${97.6}$ & ${98.0}$ & ${99.1}$ & ${99.1}$& ${98.9}$ & ${99.1}$ \\ \midrule
    Kraken & ${96.6}$ & ${96.7}$ & ${96.5}$ & ${96.6}$ & ${98.4}$ & ${98.5}$ & ${98.4}$ & ${98.5}$ \\
    Centrifuge & ${97.4}$ & ${97.4}$ & ${97.2}$ & ${97.3}$ & ${98.4}$ & ${98.6}$ & ${98.3}$ & ${98.6}$\\
    \bottomrule
  \end{tabular}
  \caption{\textbf{Precision on the Nanopore datasets} from~\citet{nicholls2018ultra} at  
  two of the finer levels in the taxonomy $\CC{T}$, genus and species. GeNet achieves high precision and is competitive with Kraken 
  and Centrifuge.}
\label{tab:prec_mock_communities}
\end{table*}

\begin{table*}
\centering
  \begin{tabular}{lSSSSSSSS}
    \toprule
    \multirow{2}{*}{\textbf{Method}} &
      \multicolumn{2}{c}{$\CC{D}_1$ (\%)} &
      \multicolumn{2}{c}{$\CC{D}_2$ (\%)} &
      \multicolumn{2}{c}{$\CC{D}_3$ (\%)} & 
      \multicolumn{2}{c}{$\CC{D}_4$ (\%) } \\
      & {Genus} & {Species} & {Genus} & {Species} & {Genus} & {Species} & {Genus} & {Species}\\
      \midrule
    GeNet & ${62.3}$ & ${39.0}$ & ${63.1}$ & ${39.5}$ & ${62.8}$ & ${31.7}$ & ${58.5}$ & ${22.0}$ \\
    GeNet top 5 & ${81.3}$ & ${68.2}$ & ${82.6}$ & ${69.6}$ & ${81.0}$ & ${73.7}$ & ${77.5}$ & ${66.3}$ \\ \midrule
    Kraken & ${94.1}$ & ${93.3}$ & ${94.0}$ & ${93.2}$ & ${96.8}$ & ${96.1}$ & ${97.1}$ & ${96.5}$ \\
    Centrifuge & ${95.2}$ & ${94.4}$ & ${95.0}$ & ${94.3}$ & ${97.5}$ & ${97.1}$ & ${97.6}$ & ${97.2}$\\
    \bottomrule
  \end{tabular}
  \caption{\textbf{Recall on the Nanopore datasets} from~\citet{nicholls2018ultra} at the  
  genus and species levels. Kraken and Centrifuge achieve 
  significantly higher recall than GeNet, due to the strong dataset shift in the 
  distribution of the labels and the noise distribution. For example, $93\%$ of $\CC{D}_3$ 
  is composed of one bacterial species only. GeNet still performs way above 
  chance ($0.13\%$ at the genus level and $0.05\%$ at the species level), 
  testifying to a significant amount of domain adaptation.}
\label{tab:recall_mock_communities}
\end{table*}
We provide experimental analysis of GeNet in two different problems. 
Section~\ref{sec:exp_da} analyses the generalization ability of GeNet to reads for which dataset shift occurs.  
Section~\ref{sec:exp_features} shows that the representations learned by GeNet are useful for downstream 
tasks, a fundamental advantage. 


We compare GeNet to two state-of-the-art methods for metagenomic classification: Centrifuge~\citep{kim2016centrifuge} 
and Kraken2~\citep{wood2014kraken}\footnote{For Centrifuge, we use the ``Bacteria, Archaea, Viruses, Human'' database 
available in \url{https://ccb.jhu.edu/software/centrifuge/}. For Kraken, we use the full standard database. 
We write Kraken instead of Kraken2.}. 
We now introduce the training dataset and training details.
The code to train and evaluate GeNet will be made available at \url{https://github.com/mrojascarulla/GeNet}. 

\begin{figure}
    \centering
    \includegraphics[width=\linewidth]{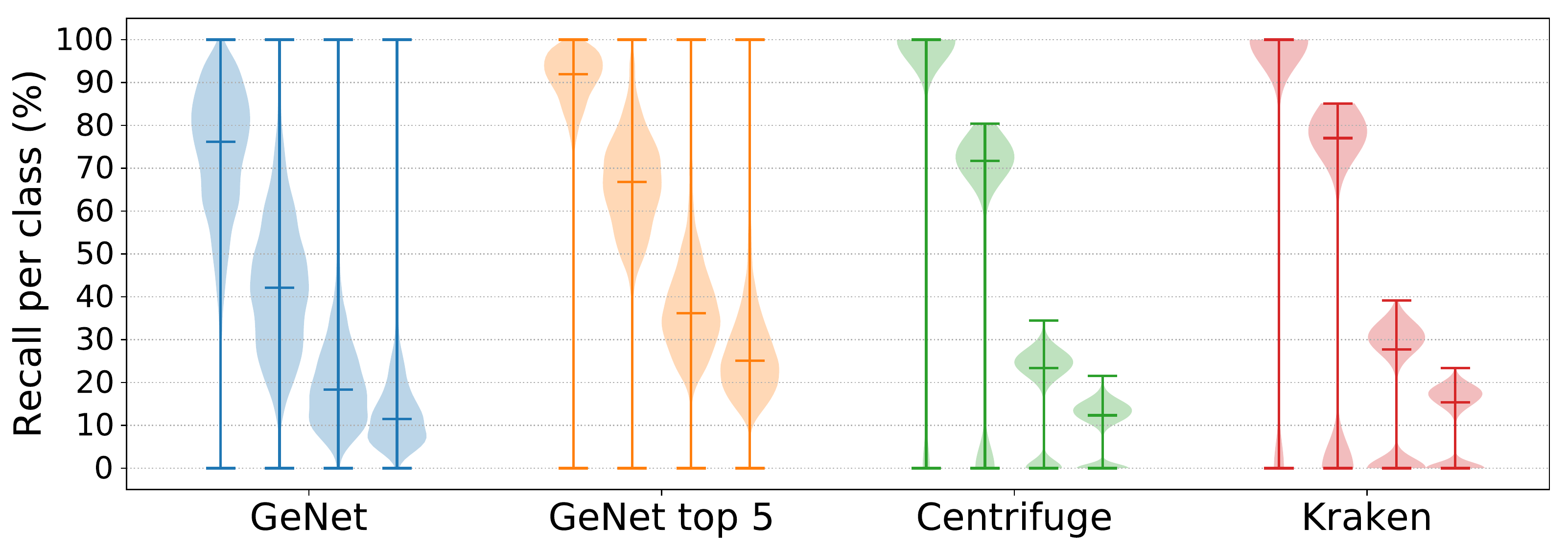}
    \caption{\textbf{Recall per class at the genus level on Illumina datasets}. For each method, four 
    results are available, corresponding to datasets with increasing read noise $q\in\{0.01, 0.1, 0.2, 0.3\}$. 
    The median recall per class is reported. The average performance of Kraken and Centrifuge 
    is higher than GeNet. For the two higher noise settings, GeNet top 5 outperforms Kraken and Centrifuge. 
    As noise increases, the distribution of per class recall 
    is bi-modal for Centrifuge and Kraken, and close to half of the labels have recall 
    near zero, which is undesirable. The distribution of per class recall for GeNet is uni-modal, 
    and remains similar as noise increases.}
    \label{fig:illumina}
\end{figure}

\begin{figure}
    \centering
    \includegraphics[width=\linewidth]{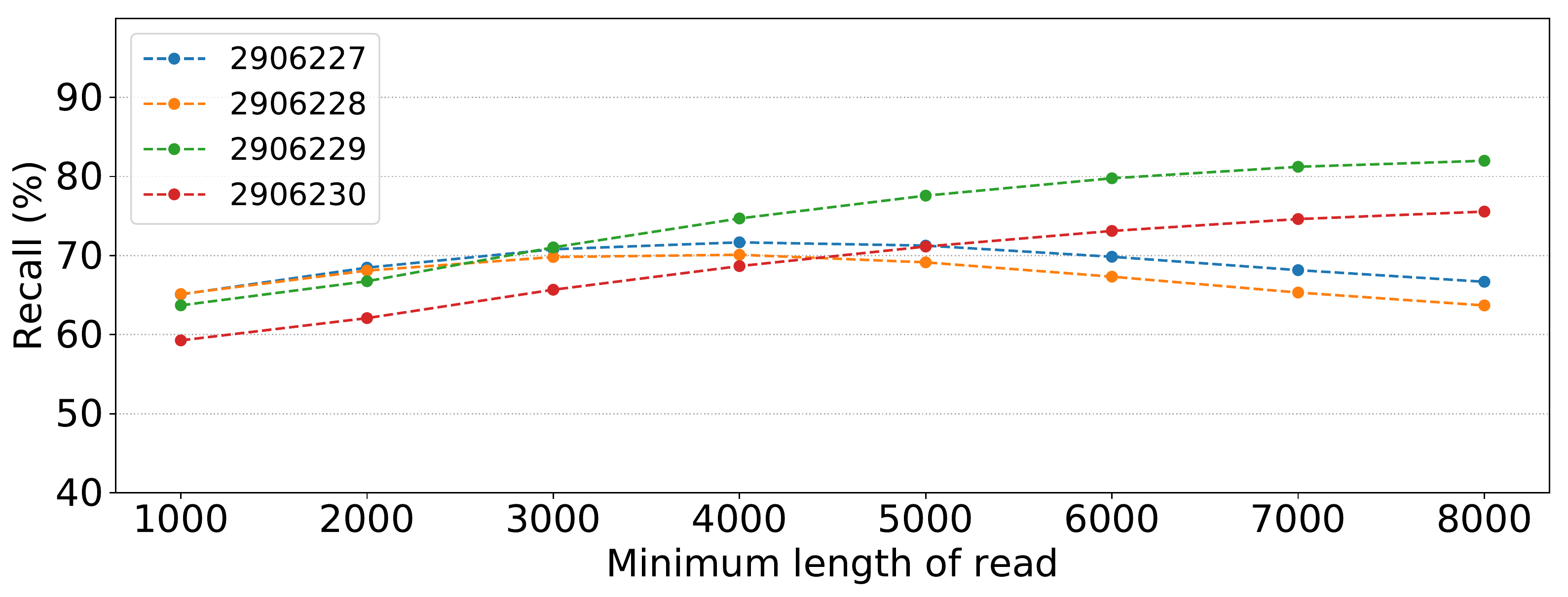}
    \caption{\textbf{Recall of GeNet at the genus level on the four Nanopore datasets} from~\citet{nicholls2018ultra} as a function of minimal read length. 
    For two datasets, GeNet performs better on longer reads, earning a $15\%$ difference between 
    all reads longer than $1,000$ to reads longer than $8,000$. Recall is roughly unchanged on the 
    other two datasets. }
    \label{fig:length}
\end{figure}

\paragraph{Training dataset and parameters.}
We use a variation of the dataset from~\citet{kim2016centrifuge} to train GeNet as described in 
Section~\ref{sec:genet}. 
This dataset consists of $4278$ prokaryotic 
genomes available in RefSeq~\citep{pruitt2013refseq}. 
To avoid redundancies, 
we removed genomes 
which shared the same taxonomy 
at the leaf level. This resulted in a dataset with $K = 3375$ genomes, 
whose NCBI reference IDs may be found at \url{https://github.com/mrojascarulla/GeNet}.  

These $K$ genomes are loaded in memory and GeNet is trained, with mini-batches generated as 
described in Algorithm~\ref{alg:sample_mb}. 
We trained i) three networks with different values of the noise parameter $p\in\{0.03, 0.1, 0.2\}$ 
for reads of fixed length $r_{max} = r_{min} = 1000$ and ii) a network with noise $p=0.1$, $r_{min} = 1000$ and 
$r_{max} = 10000$. 
For each network, we carry out 
hyper-parameter search for some network parameters using validation data generated identically to the training data, 
see Table~\ref{tab:hyperparams} in Appendix~\ref{sec:traindetails}.

We optimize the network with mini-batches of size $64$ using Nesterov momentum~\citep{sutskever2013importance}, 
with 
momentum parameter $0.9$. Training to convergence takes roughly one week on a P-100 GPU. Details on 
the architecture and hyper-parameters are provided in Appendix~\ref{sec:traindetails}. 

\paragraph{Performance metrics}
We report three performance metrics. First, we consider recall. For a dataset with 
$p$ classes, denote by $r_i$ the one-vs-all recall measured for class $i$ 
and $w_i$ the proportion of examples corresponding to class $i$ in the dataset. Then 
the overall recall is $r=\sum_{i=1}^p w_i r_i$. This is equivalent to standard accuracy. 
Similarly, we report precision $p=\sum_{i=1}^p w_i p_i$, where $p_i$ is the one-vs-all 
precision of class $i$. 

We also report the Average Nucleotide Identity (ANI), computed using FastANI~\citep{jain2018high}. 
ANI is a measure of similarity between two genomes. We report the ANI between 
the genome predicted by GeNet and the true genome, indicating if the network is making 
``sensible'' mistakes by predicting incorrect but similar labels. FastANI returns zero for 
similarity values smaller than a given threshold. In such cases, we use Average Amino-Acid 
Identity (AAI) instead, computed with CompareM~\citep{comparem}. 
Details for ANI and AAI computations are given in Appendix~\ref{sec:ani}.

Results are given both at the biological ranks (levels in the tree $\CC{T}$) of 
\emph{genus} and \emph{species}, two of the finer levels of the taxonomy, see Figure~\ref{fig:taxotree}. 
In addition to GeNet, we report GeNet top 5, in which we consider a read to be 
correctly classified if the true label is among the top 5 highest ranked predictions for 
that read.

\begin{table}
\centering
  \begin{tabular}{lc}
    \toprule
    \textbf{Method} & Accuracy \\ \midrule
    GeNet + LIN &  $\mathbf{90.3 \pm 0.1}$ \\
    Freq + LIN & $50.8 \pm 1.0$ \\
    Freq + MLP & $61.4 \pm 1.4$ \\ \midrule
    Centrifuge &$88.5$ \\
    Kraken &  $\mathbf{99.5}$\\
    \bottomrule
  \end{tabular}
  \caption{\textbf{Accuracy in the pathogen detection problem}. 
  $21$ closely related organisms are considered, $11$ of which are 
  pathogenic. 
  The logistic regression models and MLP are trained using a labeled set of size 
  $10,000$, and accuracy is computed on the remaining held-out data. Results are averaged over 
  $10$ such repetitions. 
  The recalls obtained by Kraken and Centrifuge on the whole dataset are also reported for reference. 
For pathogenesis, species level classification is necessary, so 
a prediction for these two methods is considered correct if the read is detected correctly at the species level. 
  \emph{GeNet + LIN} significantly outperform \emph{Freq + LIN} and \emph{Freq + MLP}, which exhibit 
  near chance performance. While not directly comparable, since these methods do not 
  have access to an additional training phase, \emph{GeNet + LIN} outperforms Centrifuge 
  on this task.}
\label{tab:pathogens}
\end{table}

\begin{table*}
\centering
  \begin{tabular}{lcccc}
    \toprule
    \textbf{Method} & $\CC{D}_1$(\%) & $\CC{D}_2$(\%) & $\CC{D}_3$(\%) & $\CC{D}_4$(\%) \\ \midrule
    GeNet {+ LIN}  & $\mathbf{91.1 \pm 0.0}$ & $\mathbf{92.0 \pm 0.1} $ & $\mathbf{98.6 \pm 0.0}$ &  $\mathbf{98.6 \pm 0.0}$    \\\midrule
    Freq + LIN  & $55.1 \pm 0.2$ & $55.5 \pm 0.3$ & $97.6 \pm 0.0$ &   $97.5 \pm 0.0$  \\
    Freq + MLP  & $55.7 \pm 0.2$ & $56.3 \pm 0.2$ & $97.6 \pm 0.0$ &   $97.6 \pm 0.0$ \\
    \bottomrule
  \end{tabular}
  \caption{\textbf{Recall at the species level on the four Nanopore datasets}. GeNet is used to compute 
  representations of the reads in the datasets, and a logistic regression is trained on a 
  labeled dataset with $10,000$ reads. Features built using nucleotide frequencies are  
  used as a benchmark, see Section~\ref{sec:exp_features}. Recall is computed on the remaining held-out data. Averages over $10$ repetitions are reported. 
  \emph{GeNet + LIN} significantly outperforms \emph{Freq + LIN} and \emph{Freq + MLP}, and is closer to 
  the recalls obtained by Centrifuge and Kraken. \emph{Freq + LIN} and \emph{Freq + MLP} achieve high recall on 
  two datasets, however, these datasets are highly unbalanced and are composed over $90\%$ 
  of one type of bacteria. These models always predict the majority class, and obtain zero recall for other classes. This is not 
  the case for \emph{GeNet + LIN}.}
\label{tab:genet_lin}
\end{table*}

\subsection{Generalization to other data distributions}\label{sec:exp_da}
We first evaluate the performance of GeNet on data drawn from different distributions, on 
which domain shift occurs. We consider two problems of increasing difficulty. 

First, we build a dataset with reads belonging to the $K$ genomes used during training. 
These reads are generated with added Illumina-type noise using the random-reads module 
of BBMap~\citep{bbmap}, an open-source short read aligner. For this experiment, the domain shift only 
occurs in the distribution of the noise, not the distribution of the labels. We generate 
reads with Illumina noise levels $q\in\{0.01, 0.1, 0.2, 0.3\}$. For each noise level, 
we generate ten datasets with a different random seed, 
each with one million reads split uniformly between the $K$ genomes. 
We test with GeNet trained on uniform noise parameters 
$p=0.03$ for $q=0.01$, $p=0.1$ for $q=0.1$ and $p = 0.2$ for higher $q$. 
Figure~\ref{fig:illumina} 
reports \emph{per class} recall at the genus level. Conclusions from the recall at species level are similar. 
Details on the parameters of BBMap are given in Appendix~\ref{sec:illumina}. 

\begin{figure}
    \centering
    \includegraphics[width=\linewidth]{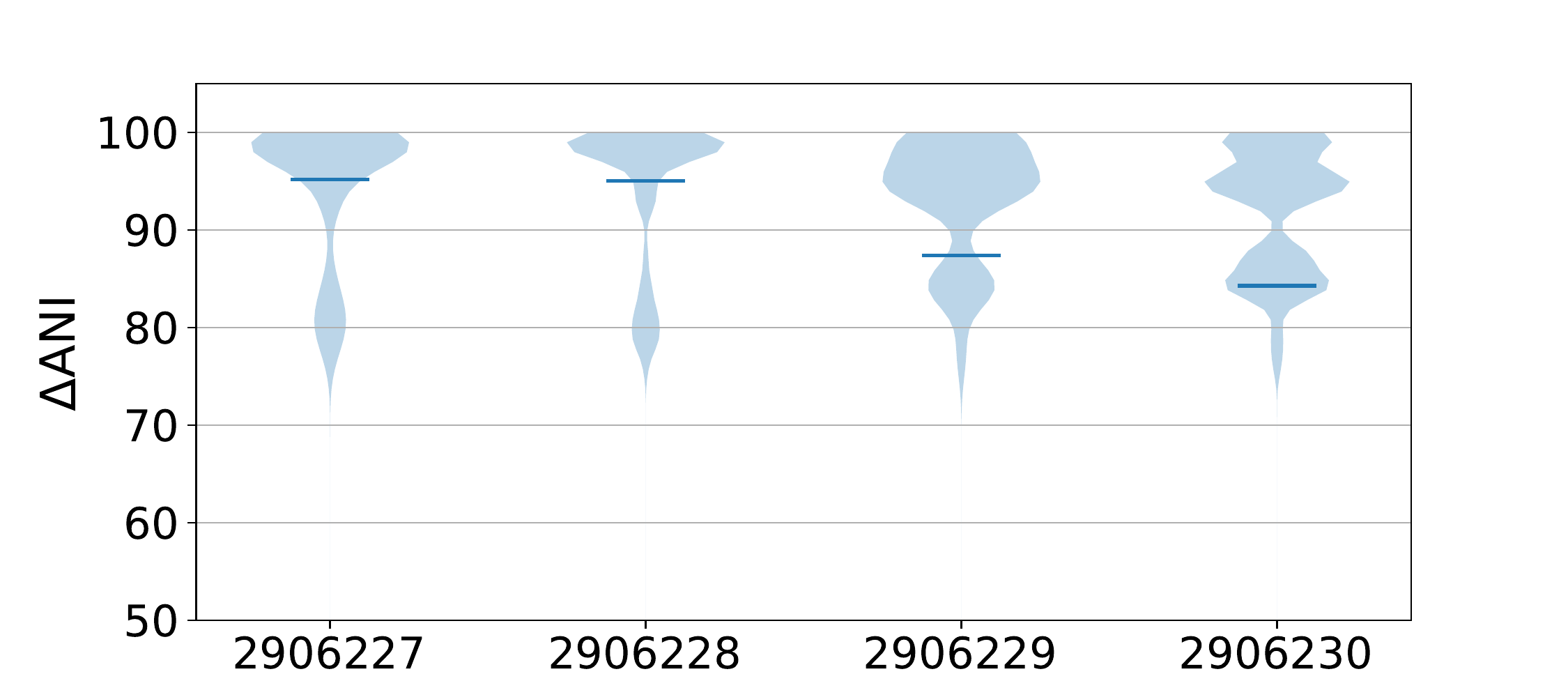}
    \caption{Average Nucleotide Identity (ANI) between genome the genome with the highest 
    predicted probability for GeNet and the true genome on the four Nanopore datasets. For values 
    of ANI smaller than $70$, Amino Acid Identity (AAI) was used instead. Since GeNet 
    was only trained with reads longer than $1,000$, we discard shorter reads in this plot, 
    excluding roughly $7\%$ of predictions in $\CC{D}_1$ and $\CC{D}_2$ 
    and $1.5\%$ on the other two datasets. For the first two datasets, median ANI is close to 
    $95\%$, and closer to $85\%$ for the other unbalanced datasets.}
    \label{fig:ani_mock}
\end{figure}

Second, we consider four real world datasets of Nanopore reads introduced in~\citet{nicholls2018ultra}, with 
the following accession numbers in the European Nucleotide Archive: \texttt{ERR2906227}, \texttt{ERR2906228}, 
\texttt{ERR2906229} and 
\texttt{ERR2906230}. Below, we call these datasets $\CC{D}_1, \ldots,  \CC{D}_4$. 
Both $\CC{D}_1$ and $\CC{D}_3$ contain around $3$ million reads, while $\CC{D}_2$ and $\CC{D}_4$ 
contain over $30$ million reads. 
These datasets contain reads from $10$ organisms, of which we only consider $8$ (the remaining 
two eukaryotes, Cryptococcus neoformans and Saccharomyces cerevisiae, 
were not observed during training, so GeNet cannot classify them into the correct 
class). To obtain ground truth labels, we map the reads to the 
genomes present in the dataset using minimap2~\citep{li2018minimap2}, a DNA aligner. Details 
on the use of minimap2 are given in Appendix~\ref{sec:minimap}. 
We report precision and recall in Tables~\ref{tab:prec_mock_communities} and~\ref{tab:recall_mock_communities} 
respectively. Moreover, we report Average Nucleotide Identity (ANI) between the predictions of GeNet 
and the true labels in Figure~\ref{fig:ani_mock}. We also analyse the effect of read length in 
accuracy in Figure~\ref{fig:length}.

\subsection{Downstream tasks with learned representations}\label{sec:exp_features}
We showcase the re-usability of the representations in the last hidden layer of GeNet in two downstream 
tasks as described in Section~\ref{sec:da}. 
Given a labelled dataset $\CC{D} = \{(\B{s}_i, z_i)\}_{i=1}^n$, we train a linear 
logistic regression on the transformed dataset $\widetilde{\CC{D}} = \{(\B{h}_i, z_i)\}_{i=1}^n$, 
where the label $z_i$ depends on the specific example 
and $\B{h}\in\BB{R}^{1024}$. 
We denote this 
method \emph{GeNet + LIN}. As an alternative representation of the reads, 
we divide a sequence $\B{s}_i$ in ten equally sized 
bins, and compute the frequency of each of the four nucleotides in every bin. The resulting $40$ dimensional 
vector $\widehat{\B{h}}_i$ is an alternative representation of $\B{s}_i$. 
As baselines, we train a logistic regression and a Multi Layer Perceptron on the transformed dataset 
$\widehat{\CC{D}} = \{(\widehat{\B{h}}_i, z_i)\}_{i=1}^n$ with grid search and cross validation, 
resulting in methods \emph{Freq + LIN} and 
\emph{Freq + MLP}. 

As a first downstream task, we consider the four Nanopore datasets discussed in 
Section~\ref{sec:exp_da}. We sub-sample each dataset to contain only $3$ million reads 
for computational reasons. In each dataset separately, we use $10,000$ randomly chosen 
examples with 
their true labels \emph{at the species level} 
as a labelled dataset $\CC{D} = \{(\B{s}_i, y_i)\}_{i=1}^n$. Note 
that this is only $0.33\%$ of the size of each dataset. We evaluate performance on the 
remaining held-out reads and average the results over ten repetitions, see Table~\ref{tab:genet_lin}.

As a second task, we build a dataset of $21$ closely related genomes, of which $10$ are 
highly pathogenic (harmful) and the rest are not known to cause any diseases. The dataset 
consists of 
$4$ Clostridium strains, one Clostridiodes strain, $4$ Vibrio strains, $5$ Pseudomonas 
  strains, $2$ Klebsiella strains, $3$ Streptococcus strains and $2$ Burkholderia strains (each of these 
  indicates a different genus). Within each genus, only some of the strains are pathogenic. 
  The corresponding NCBI 
accession IDs can be found at \url{https://github.com/mrojascarulla/GeNet}. 
We consider the labelled dataset 
$\CC{D} = \{(\B{s}_i, z_i)\}_{i=1}^n$ where $z_i$ is a binary label indicating whether the 
corresponding read belongs to a pathogenic genome. The reads are generated using BBMap, with 
PacBio noise of $10\%$. The dataset has $100,000$ reads, of which $10,000$ randomly chosen 
reads are used for training. Accuracy is evaluated on the remaining held-out reads, averaged 
over $10$ repetitions, see Table~\ref{tab:pathogens}.

\begin{table}[th]
\centering
\begin{tabular}{lcc}
\toprule
\textbf{Method} & Speed (reads/min)  & Memory (GB) \\ \midrule    
GeNet, P-100 & $71,000$ & $\mathbf{0.126}$  \\ 
GeNet, P-40 & $57,000$ & -  \\\midrule
Kraken & $\mathbf{1,732,000}$  & $93.0$ \\
Centrifuge & $563,380$ &  $11.3$ \\ \bottomrule
\end{tabular}
\caption{Speed and memory requirements of metagenomic classifiers. Speed is computed during 
the evaluation of dataset $\CC{D}_3$, which contains over $3$ million reads. 
We report speed for GeNet using both P-40 and P-100 GPU cards. 
Kraken uses $8$ CPU cores and 
$120$ GB of RAM. Speed for Centrifuge is reported from~\citet{kim2016centrifuge}, 
since Centrifuge does not return how much time was spent loading the database and how much 
classifying the reads. GeNet is slower than the competitors, inference 
is only an order of magnitude slower than for Centrifuge. However, 
the memory usage of GeNet is smaller 
by two orders of magnitude.}
\label{tab:speed_memory}
\end{table}
\subsection{Analysis of the results}
We draw the following conclusions from our experiments. 

\paragraph{GeNet partially generalises to real data.}
Our experiments show that despite the occurring dataset shift, GeNet can perform significantly 
above chance in real datasets at fine levels of the hierarchical taxonomy. 
For Illumina reads, the recall achieved by Kraken and Centrifuge is 
higher than that of GeNet. Nonetheless, Figure~\ref{fig:illumina} shows that as noise increases, 
Kraken and Centrifuge achieve recall close to zero for a large proportion of the genomes. 
A more uniform recall distribution per genome may be desirable, and is partially achieved by GeNet for which 
few classes have recall close to zero, even as noise increases. This showcases the usefulness 
of training with noise. 

On the Nanopore datasets, recall for GeNet 
is significantly lower than for Centrifuge and Kraken, see Table~\ref{tab:recall_mock_communities}, while precision is competitive, see Table~\ref{tab:prec_mock_communities}. We attribute the low recall mainly to the 
strong distribution shift, both in the output labels (only $8$ genomes are present in the datasets, 
while we used $3375$ for training, and $\CC{D}_3$ 
and $\CC{D}_4$ are highly unbalanced) and 
the noise distribution. Nonetheless, the ANI distribution in Figure~\ref{fig:ani_mock} 
witnesses that often the predicted genomes are similar to the 
true genomes, which means that most mistakes are not unreasonable. Moreover, 
GeNet is expected to perform better as read length increases, which is the case in 
two of the datasets, see Figure~\ref{fig:length}.

\paragraph{GeNet representations perform well in downstream tasks.}
Experiments on downstream tasks show that the representations learned by 
GeNet can be successfully exploited. While GeNet significantly under-performs state-of-the-art 
methods in terms of recall on the Nanopore datasets, training a linear model on top 
of GeNet representations leads to a significant increase in performance in these datasets, 
see Table~\ref{tab:genet_lin}. While 
we cannot compare the results directly with Centrifuge and Kraken since these do not have 
access to an extra training phase, the obtained recalls are competitive. 
In practice, a small percentage of a target dataset of reads can be labeled with alternative methods to 
train such a supervised model. 

Second, using standard frequency features on the pathogen dataset leads to close to chance 
performance. This showcases the difficulty of this problem from raw data. \emph{GeNet + LIN} achieves 
over $90\%$ held-out accuracy, see Table~\ref{tab:pathogens}. 
While direct comparison with Kraken and Centrifuge is not possible in this case, 
it is encouraging that 
\emph{GeNet + LIN} outperforms Centrifuge in this problem. 

\paragraph{GeNet strikes a trade-off between speed and storage.}
Computing predictions for reads of size $10,000$ on a NVIDIA P-100 is roughly $10$ times slower than 
Centrifuge and $20$ times slower than Kraken, see Table~\ref{tab:speed_memory}. 
Nonetheless, the only storage required by GeNet are the weights of the network, which 
require $126$MB. This is two orders of magnitude smaller than 
Centrifuge, and three orders of magnitude smaller than Kraken.

\section{Conclusion}
We showed in two datasets obtained from different sequencing technologies 
that GeNet achieves good recall and high precision despite strong 
dataset shift. 
We also provided evidence that the representations 
in the last hidden layer of GeNet can be used for downstream 
tasks. Training a linear model with a small percentage of labelled  
data in Nanopore datasets leads to an increase of recall at the 
species level of $50\%$ or more. Moreover, GeNet features significantly 
outperform frequency features computed from the raw sequences on a challenging 
pathogen 
detection problem. 

We expect that GeNet representations can be used 
for a variety of tasks in computational biology, e.g.,
gene function prediction. 
Many tasks require a \emph{representation} of the data 
and thus cannot benefit from methods such as Kraken and Centrifuge, 
notwithstanding the excellent performance these custom tools exhibit in metagenomic classification.  
In addition, our approach exhibits a higher level of noise robustness, 
the ability to learn from technology specific noise, as well as small memory requirements.
This can present interesting opportunities 
for the development of cheaper and portable sequencing technologies.

The use of pre-trained networks and data representations has accelerated research in computer vision, 
speech and natural language processing, allowing fast deployment of solutions for new problems that often come with small labelled training sets. We anticipate a similar potential 
for computational biology and health, where labelled data sets can also he hard to come by. 
The ultimate promise and validation of the proposed method would thus consist of its
adoption by the community and application in a diverse array of tasks, 
which is well beyond the scope of the present work.

\section*{Acknowledgements}
The authors are thankful to Sylvain Gelly for helpful discussions on the 
architecture design for GeNet.

\nocite{langley00}

\bibliography{refs}
\bibliographystyle{icml2019}

\clearpage
\appendix

\begin{center}
\textbf{Appendix to ``GeNet: Deep Representations for Metagenomics''}
\end{center}

\section{Training details}\label{sec:traindetails}
GeNet was implemented using Tensorflow~\citep{tensorflow2015-whitepaper}. 

\begin{figure*}
\centering
\tikzstyle{arrow} = [thick,->,>=stealth]
\tikzstyle{input} = [rectangle, rounded corners, minimum width=5cm, minimum height=1cm,text centered, draw=black, fill=black!10, align=center]
\tikzstyle{layer} = [rectangle, rounded corners, minimum width=5cm, minimum height=1.8cm,text centered, draw=black, fill=blue!10, align=center]
\tikzstyle{output} = [rectangle, rounded corners, minimum width=3cm, minimum height=1cm,text centered, draw=black, fill=red!20, align=center]
\begin{tikzpicture}[node distance=2cm]
\node (input) [input] {\Large $\B{s} \in \{0, 1, 2, 3\}^{r_{max}}$};
\node (emb) [layer, below=1.2 of input] {\Large \texttt{Embed} + \texttt{Pos.}\texttt{embed}};
\node (resnet) [layer, below=1.2 of emb] {\Large \texttt{Resnet blocks}};
\node (fc) [layer, below=1.2cm of resnet] {\Large $\B{h} = R(\B{s})$};
\node (anchor) [below=-0.15cm of fc] {};
\node (s1) [output, below=3cm of fc] {\Large  $\widehat{\B{y}}_2\in\BB{R}^{N_2}$};
\node (s2) [output, left=2cm of s1] {\Large $\widehat{\B{y}}_1\in\BB{R}^{N_1}$};
\node (s3) [output, right=2cm of s1] {\Large $\widehat{\B{y}}_3\in\BB{R}^{N_3}$};
\node (text) [above right=1cm and -0.1cm of s2] {\Large $W_1$};
\node (text) [above left=1cm and 0.1cm of s3] {\Large $W_3$};
\node (text) [above right=1cm and -1.4cm of s1] {\Large $W_2$};

\node (text) [below right=-0.4cm and .65cm of s2] {\Large $U_1$};
\node (text) [below right=-0.4cm and .65cm of s1] {\Large $U_2$};

\draw [arrow] (input) -- (emb);
\draw [arrow] (emb) -- (resnet);
\draw [arrow] (resnet) -- (fc);
\draw [arrow] (anchor) -- (s1);
\draw [arrow] (anchor) -- (s2);
\draw [arrow] (anchor) -- (s3);
\draw [arrow] (s2) -- (s1);
\draw [arrow] (s1) -- (s3);
\end{tikzpicture}
\caption{GeNet architecture. The hidden representation $\B{h}$ is mapped to 
$L$ softmax layers, representing the unnormalised probability vector over taxa at $L$ 
levels in a hierarchical taxonomy. Here, only $3$ levels of the taxonomy are 
depicted, GeNet is trained with $L=7$ taxonomic levels.}\label{fig:nn}
\end{figure*}
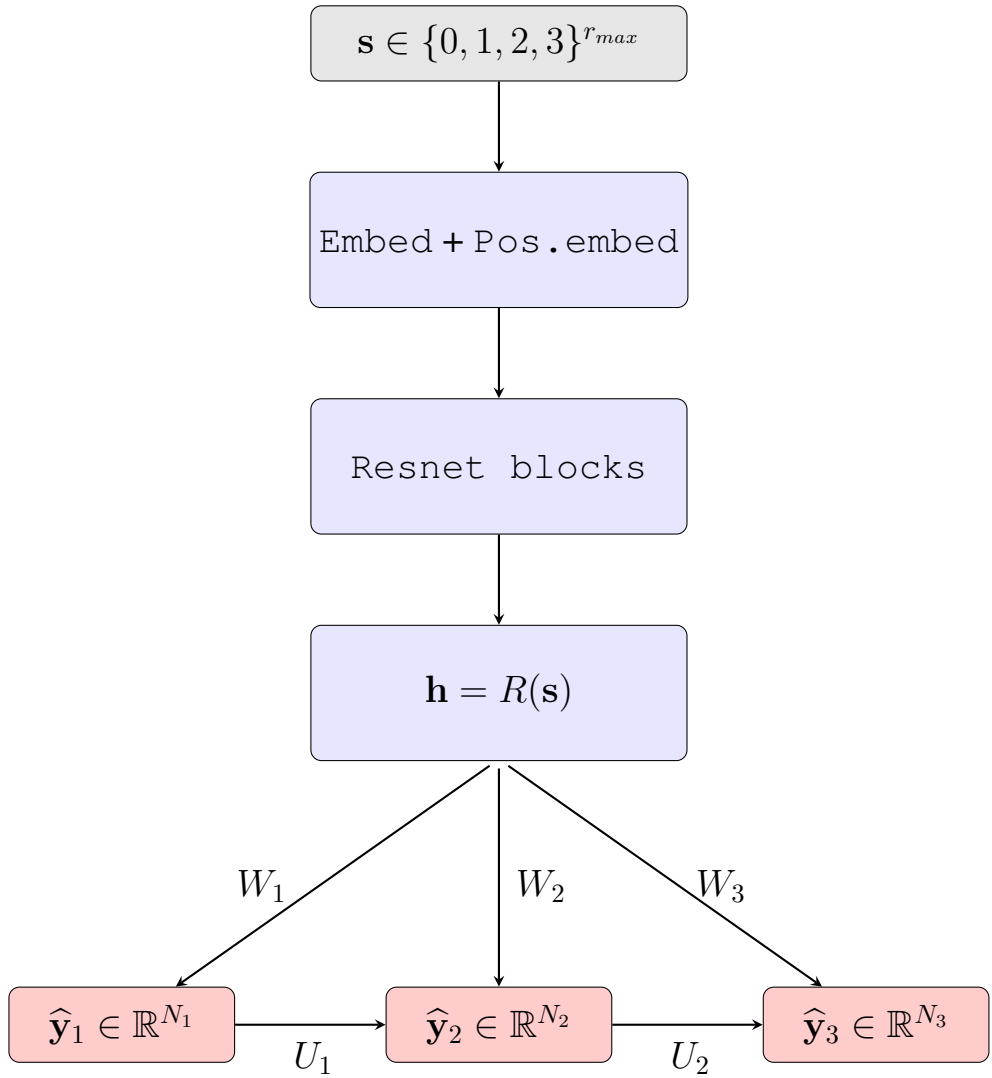

The vocabulary size is $6$: four nucleotides \texttt{A}, \texttt{C}, \texttt{T} and \texttt{G}, 
an end-of-sequence character, and a character for \emph{ambiguous} nucleotides, which appears 
occasionally on the downloaded genomes. 
The input to GeNet is a sum of the one-hot encoding of each letter in the input sequence, 
a trainable six dimensional embedding, and a trainable six dimensional positional embedding 
as proposed in~\citet{convseq2seq17}. For an input sequence of length $r_{max}$, this results 
in a matrix of shape $6 \times r_{max}$. The architecture of GeNet is available in Table~\ref{tab:arch} 
and is depicted in Figure~\ref{fig:nn}. 
Details for the Resnet blocks used can be found in Table~\ref{tab:resnetblock}. 
For every call of a Resnet block of the form $(n, 2n)$, the number of filters is multiplied by two, 
and the size of the input is divided by two. When there is a size mismatch between the 
input to the Resnet block and the output, a $1$d convolution with the appropriate number of 
filters is used to match the dimensions. 

We perform grid search for some of the hyper-parameters of the network, the ranges considered are in 
Table~\ref{tab:hyperparams}. We selected the final version of GeNet based purely on validation accuracy 
on data drawn from the same distribution as the training data.

\begin{table}[h]
\begin{center}
\begin{tabular}{c}
      \multicolumn{1}{c}{\textbf{Resnet block} $n\times 2n$}\\
      \toprule
      \textbf{Layers} \\ \midrule
       \texttt{AVG. POOL $1\times 2$, \texttt{stride} $2$}. \\
       \texttt{BN + ReLU}. \\
       \texttt{CONV} $1\times w \times 2n$, \texttt{stride} $1$. \\
       \texttt{BN + ReLU}. \\
       \texttt{CONV} $1\times w \times 2n$, \texttt{stride} $1$. \\
       \texttt{Input + CONV} \\
      \bottomrule
    \end{tabular}
\end{center}
\caption{Resnet block used in GeNet. The input is added to \texttt{CONV} 
in the last layer using a $1\times 1$ convolution with $2n$ filters. 
Resnet blocks of the form $(n, n)$ do not perform average pooling as a first stage, 
so the input and output number of filters is unchanged. }
\label{tab:resnetblock}
\end{table}

\begin{table}[h]
\begin{center}
\begin{tabular}{c}
      \multicolumn{1}{c}{\textbf{GeNet}}\\
      \toprule
      \textbf{Layers} \\ \midrule
       \texttt{Embedding + Pos.embedding + One-hot.} \\
       \texttt{CONV} $v \times w \times n_f$, \texttt{stride} $w$. \\
       \texttt{RESNET BLOCK $n_f \times n_f$} \\
       \texttt{RESNET BLOCK $n_f \times n_f$} \\
       \texttt{RESNET BLOCK $n_f \times 2n_f$} \\
       \texttt{RESNET BLOCK $2n_f \times 2n_f$} \\
       \texttt{BN + ReLU} \\
       \texttt{AVG. POOL + BN} \\
       \texttt{FC} $f_c$.\\
       \texttt{L softmax layers}.\\
      \bottomrule
    \end{tabular}
\end{center}
\caption{Layers of GeNet. \texttt{BN} stands for Batch Norm~\citep{bn2014}, 
\texttt{FC} for fully connected. \texttt{CONV} $v\times w\times n_f$ stands for a 
2d convolutional 
layer with kernel size $(v, w)$ and $n_f$ filters. }\label{tab:arch}
\end{table}

\begin{table}
\centering
  \begin{tabular}{lc}
    \toprule
    Parameter &  Values \\
      \midrule
    Number of initial filters $n_f$ & $\{128,256\}$\\
    Size of 1d kernel $v$    & $\{2, 3, 5\}$ \\
    Learning rate $l_r$       &  $\{0.5, 1, 10, 20\}$ \\
    Size of fully connected layer $f_c$ & $\{512, 1024\}$ \\
    Batch size $M$   &  $64$ \\ 
    \bottomrule
  \end{tabular}
  \caption{Hyperparameters for GeNet.}
\label{tab:hyperparams}
\end{table}
The model used for the experiments, which led to the highest validation accuracy, has the following parameters:
\begin{align}
    n_f &= 128 \nonumber \\ 
    v &= 3 \nonumber \\
    l_r &= 1 \nonumber \\
    f_c &= 1024 \nonumber
\end{align}

\section{Details for biological software}\label{sec:biomodules}

\subsection{Generation of Illumina and Pacbio reads}\label{sec:illumina}
Illumina reads were generated for the $K$ genomes used during training with the \texttt{RandomReads} module of 
BBMap, downloaded from \url{https://sourceforge.net/projects/bbmap}. 

For Illumina reads, we considered four noise parameters: $q\in\{0.01, 0.1, 0.2, 0.3\}$. We compute the corresponding  
Phred quality scores $n = -10\log_{10}(q)$ which are given as input to the \texttt{RandomReads} module. 

For each noise value, we generate $10$ datasets with different random seeds. Each dataset contains 
$nc =N / K$ reads of length $1,000$ from each of the $K$ genomes observed during training, 
for $N = 1,000,000$. A typical call of the module 
would look as follows:

 \texttt{randomreads.sh ref=NC\_017449.1.fasta out=NC\_017449.1.fastq len=1000 metagenome=f addpairnum=t reads=nc q=q seed=dataset\_num}

For the pathogen experiment in Section~\ref{sec:exp_features}, we generate one dataset with PacBio reads from $21$ 
organisms. A typical command would look as follows: 

 \texttt{randomreads.sh ref=NC\_017449.1.fasta out=NC\_017449.1.fastq minlength=1000 maxlength=10000 pacbio=t pbmin=q pbmax=q metagenome=f addpairnum=t reads=nc seed=-1}

\subsection{Obtain ground truth labels with minimap2}\label{sec:minimap}
We use minimap2(\url{https://github.com/lh3/minimap2}) to align the reads from the four Nanopore datasets in \citet{nicholls2018ultra} 
to the reference genomes, provided by assembling the reads in these communities using Illumina technologies. The file 
\texttt{Zymo-Isolates-SPAdes-Illumina.fasta.gz} is also provided by~\citet{nicholls2018ultra}. Given a FASTQ file with reads \texttt{reads.fastq}, 
we used minimap2 with the following paramters:

\texttt{minimap2 -ax -map-ont -t 32 Zymo-Isolates-SPAdes-Illumina.fasta.gz reads.fastq > align.sam}

\subsection{Computation of Average Nucleotide Identity (ANI) and Average Amino-Acid Identity (AAI)}\label{sec:ani}

We compute ANI using FastANI (\url{https://github.com/ParBLiSS/FastANI}) 
and AAI using CompareM (\url{https://github.com/dparks1134/CompareM}). 

Given a file \texttt{genomes.txt} containing a list of paths to the $K$ genomes is the dataset, we ran the following command:

\texttt{fastANI -rl genomes.txt -ql genomes.txt -t 8 -o similarity.out}.

This returns a value of zero for many pairs of genomes, since fastANI return zero for values under $70$. 
We completed the similarity matrix using AAI, computed with compareM as follows:

\texttt{comparem aai\_wf --cpus 32 genomes.txt aai}.


\end{document}